\documentclass[prb,reprint,superscriptaddress,super,superbib,amsmath,aps]{revtex4-1}
\usepackage{graphicx}
\usepackage[squaren]{SIunits}
\usepackage{color}
\usepackage{ulem}

\def\braket#1{{\langle{#1}\rangle}}

\begin{document}  

\title{Photon-Statistics Excitation Spectroscopy of a Single Two Level System}

\author{Max Strau\ss} 
\affiliation{Institut f\"ur Festk\"orperphysik, Quantum Devices Group, Technische Universit\"at Berlin,
Hardenbergstra{\ss}e 36, EW 5-3, 10623 Berlin, Germany}
\author{Marlon Placke}
\affiliation{Institut f\"ur Festk\"orperphysik, Quantum Devices Group, Technische Universit\"at Berlin,
Hardenbergstra{\ss}e 36, EW 5-3, 10623 Berlin, Germany}
\author{S\"oren Kreinberg}
\affiliation{Institut f\"ur Festk\"orperphysik, Quantum Devices Group, Technische Universit\"at Berlin,
Hardenbergstra{\ss}e 36, EW 5-3, 10623 Berlin, Germany}
\author{Christian Schneider}
\affiliation{Technische Physik, Physikalisches Institut, Wilhelm Conrad
R\"ontgen Center for Complex Material Systems, Universit\"at W\"urzburg, 97074
W\"urzburg, Germany}
\author{Martin Kamp}
\affiliation{Technische Physik, Physikalisches Institut, Wilhelm Conrad
R\"ontgen Center for Complex Material Systems, Universit\"at W\"urzburg, 97074
W\"urzburg, Germany}
\author{Sven H\"ofling}
\affiliation{Technische Physik, Physikalisches Institut, Wilhelm Conrad
R\"ontgen Center for Complex Material Systems, Universit\"at W\"urzburg, 97074
W\"urzburg, Germany}
\affiliation{SUPA, School of Physics and Astronomy, University of St. Andrews,
St. Andrews KY16 9SS, United Kingdom}

\author{Janik Wolters}
\affiliation{Institut f\"ur Festk\"orperphysik, Quantum Devices Group,
Technische Universit\"at Berlin, Hardenbergstra{\ss}e 36, EW 5-3, 10623 Berlin, Germany}
\affiliation{Present address: Department of Physics, University of Basel,
Klingelbergstra{\ss}e 82, CH-4056 Basel, Switzerland}
\author{Stephan Reitzenstein }
\affiliation{Institut f\"ur Festk\"orperphysik, Quantum Devices Group, Technische Universit\"at Berlin,
Hardenbergstra{\ss}e 36, EW 5-3, 10623 Berlin, Germany}

\begin{abstract}
We investigate the influence of the photon statistics on the excitation dynamics of a single
two level system. A single semiconductor quantum dot represents the two level
system and is resonantly excited either with coherent laser light, or excited
with chaotic light, with photon statistics corresponding to that of thermal
radiation. Experimentally, we observe a reduced absorption cross-section under
chaotic excitation in the steady-state. In the transient regime, the Rabi
oscillations observable under coherent excitation disappear under chaotic
excitation. Likewise, in the emission spectrum the well-known Mollow triplet,
which we observe under coherent drive, 
disappears under chaotic excitation. Our observations are fully consistent
with theoretical predictions based on the semi-classical Bloch equation
approach.
\end{abstract}

\maketitle  

The fermionic two level system (TLS) is the prototype of a quantum system. 
As a realization of the quantum bit it finds a plethora of applications in
quantum information
processing\,\cite{Devoret2013a,0953-8984-18-21-S08,Nielsen2010}.
Hence it is not surprising that two level systems under coherent excitation, e.g. under excitation 
with laser light or microwaves, are vastly studied and constitute a principal
topic in any textbook on quantum physics.  
Today, the interaction of individual TLSs with coherent radiation or even single photons is routinely studied 
in many experiments with single atoms and ions in the gas phase, defect centers in wide band-gap materials
or semiconductor quantum dots\,\cite{Leibfried2003,Meyer2015,Wolters2013a}.
These experiments form the basis of many exciting applications in quantum technology.
Interestingly, while the case of non-classical excitation statistics has been
studied in various works \cite{Georgiades1995,Murch2013}, the influence of thermal excitation statistics 
on single TLSs still needs to
 be experimentally explored, in atomic
 as well as in solid state systems.
The underlying physics of this open question is of great interest from a fundamental point of
view and is also motivated by the fact that coherent excitation conditions are rather artificial as virtually all
 radiation occurring in nature, e.g.  black-body radiation or bremsstrahlung, is
of chaotic nature.\\
In this Rapid Communication we set out to experimentally investigate
the resonant excitation of single semiconductor quantum dots in
the so far unexplored regime of resonant driving with chaotic light. 
In our comprehensive studies we compare fluorescence intensity, emission spectra 
and dynamics of a two level system represented by a semiconductor quantum
dot (QD) under excitation with coherent and chaotic light.
In the steady-state, we find a reduced absorption cross-section under chaotic
excitation.
% and at the
%same time an unaltered absorption spectrum. - ?? STIMMT DAS ?? - 
In addition, in the emission spectrum the well-known Mollow triplet present
under coherent drive of the TLS disappears under chaotic excitation, and likewise no signatures of Rabi oscillations
are observed in the time domain.
At the same time, the non-classical character of the photon emission of the TLS
is preserved under chaotic excitation, as shown in second order auto-correlation measurements.
All of these experimental findings are in excellent agreement with a
quantum mechanical description of the experimental condition.

Coherent light exhibits a Poissonian photon number distribution, where the
probability $p_{\textrm{pd}}(n)$ to observe a certain photon number $n$ is given
by
\begin{eqnarray}
p_{\textrm{pd}}(n)&=&\frac{\braket{\hat n}^{n}\exp(-\braket{\hat n})}{n!},
\end{eqnarray}
where the mean photon number is $\braket{\hat n}=\braket{\hat a^{\dag}\hat a}$,
with the conventional creation (annihilation) operator $\hat a^{\dag}$ ($\hat a$). 
The standard deviation of the photon number is given by $\Delta
n=\sqrt{\braket{\hat n}}$ and for
 large $\braket{\hat n}$ the intensity fluctuations become negligible.
Assuming stationarity, the second order auto-correlation function   
\begin{eqnarray}
g^{(2)}(\tau)&=&\frac{\braket{:\hat n(0)\hat n(\tau):}}{\braket{\hat n(0)}^{2}},
\end{eqnarray}
with $:...:$ indicating normal ordering, is constant $g^{(2)}(\tau)=1$.
Conventional laser radiation well above the laser threshold is a very accurate
realization of such coherent light. 
\begin{figure}[htbp]
\center  
\includegraphics[width=0.5\textwidth]{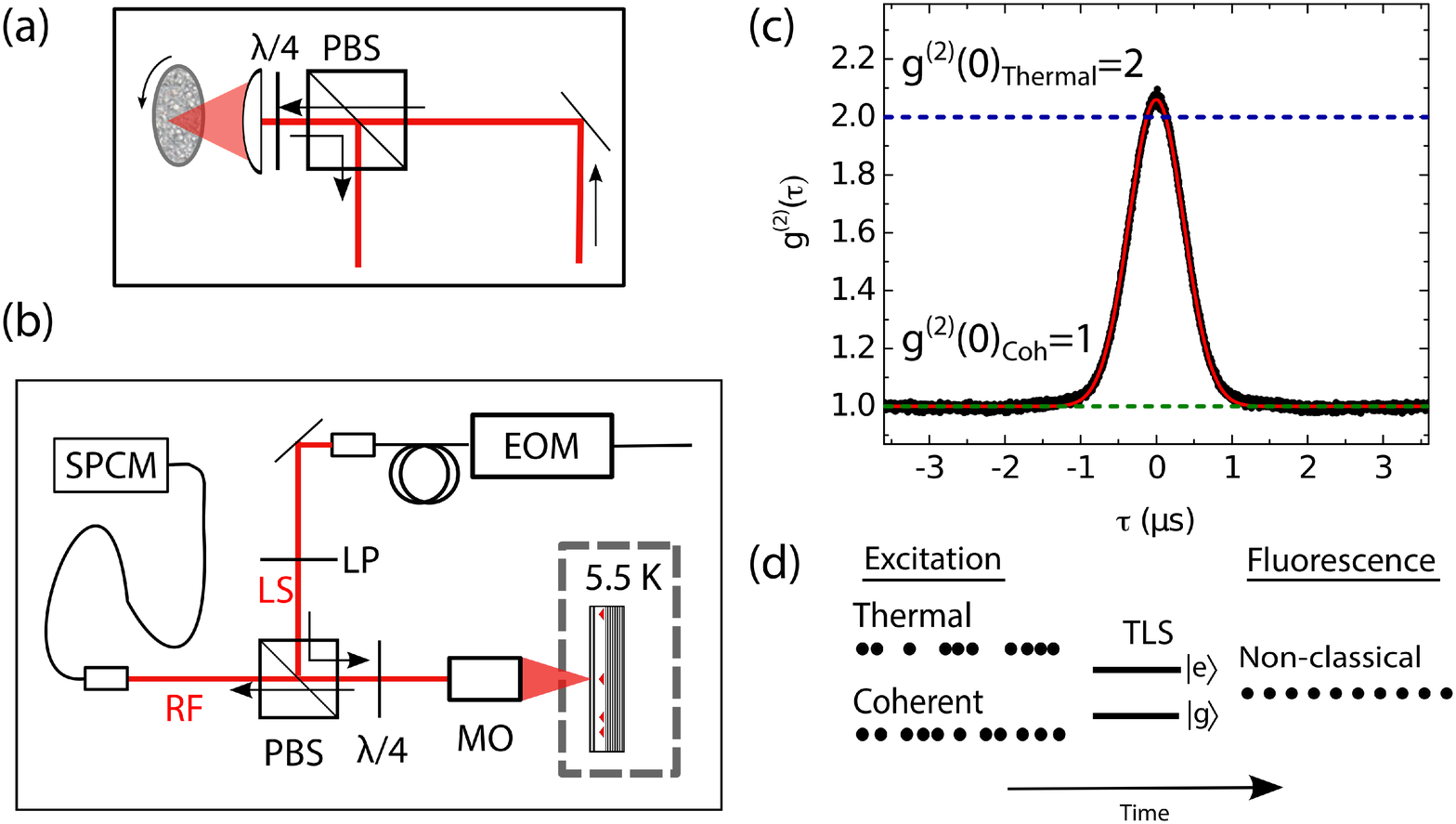}  
\caption{(Color online) (a) Sketch of the implemented Martienssen lamp.
(b) Sketch of experimental setup: Polarization filtering
used to distinguish resonant fluorescence (RF) from light source (LS).  
 Akronyms used in sketch: 
EOM: electro optical modulator,
% FBS \textit{fiber based beam splitter}, 
LP: linear polariser,
MO: microscope objective
PBS: polarising beam splitter
%, RE \textit{resonant emission} 
SPCM: single photon counting module   
(c) Measured $g^{(2)}(\tau)$-function of the chaotic light source.
The red, solid line is a fit of
$f(\tau)=~1~+~A~\textrm{exp}[-\pi(\frac{\tau}{\tau_{\textrm{corr}}})^2]$ to
the data giving a correlation time of 0.9\,$\mu$s. (d) Simplified picture of
experiment illustrating different photon statistics involved in the experiment.}
\label{fig:aufbau}
\end{figure} 
  
In contrast, chaotic light follows the Bose-Einstein statistics and the
probability $p_{\textrm{ch}}(n)$ to observe a certain photon number $n$ is given
by
\begin{eqnarray}
p_{\textrm{ch}}(n)&=&\frac{\braket{\hat n}^{n}}{(1+\braket{\hat
n})^{n+1}}.\label{eq:chaotic_distribution}
\end{eqnarray}
The  fluctuations of the photon number are given by $\Delta
n=\sqrt{\braket{\hat n}+\braket{\hat n}^{2}}$. For large 
$\braket{\hat n}$ the fluctuations of the photon number are on the order of the average 
photon number $\braket{\hat n}$. Assuming stationarity, the Fourier transform of the spectrum
$g^{(1)}(\tau)=\braket{\hat a^{\dag}(0)\hat a(\tau)}/\braket{\hat n}$ of a
chaotic light field determines its second order auto-correlation function:
\begin{eqnarray}
g^{(2)}(\tau)&=&1+|g^{(1)}(\tau)|^{2}.
\end{eqnarray}
Obviously  $g^{(2)}(0) = 2$, i.e. chaotic light shows photon bunching leading to considerable effects
 in non-linear spectroscopy, e.g. an enhanced two-photon
 absorption probability\,\cite{Carmele2009,Kazimierczuk2015,Boitier2009}.
Ideal black-body radiation, or emission from an infinite number of independent emitters are natural sources 
of chaotic light\,\cite{Loudon2001}. 
However, as these sources have limited spectral brightness and \'etendue, their
use in nonlinear spectroscopy is very restricted.
To circumvent these limitations, we implement a chaotic light source with a
Gaussian spectrum, also known as Martienssen 
lamp\,\cite{Martienssen1964,Arecchi1965, Martienssen1966}, by reflecting a
focused laser beam on a circular diffuser (1500\,Grit) moving with a constant
velocity of $v \approx  10$\,m/s at a radius of 10\,mm (see Fig.\,1\,(a)).
The diffuse reflection on the multitude of moving scatterers introduces Doppler
broadening of the spectrum and chaotic intensity fluctuations.
Fig.\,1\,(c) shows the measured second-order autocorrelation function of the
used source, exhibiting a second-order correlation time of $\tau^{(2)}_{\textrm{corr}} =
(901.8 \pm 0.9)$\,ns according to a Gaussian fit (red trace) of the correlation
function. The correlation time of the thermal field can be altered by adjusting
the angular frequency of the diffuser. Furthermore, the $g^{(2)}(0)$ value of 2.05 shows that the
source produces light with almost perfect thermal statistics at an emission linewidth of 1.1 MHz,
where the slight deviations can be attributed to mechanical instabilities of the setup leading to an increased
bunching. 

As TLS we use single self assembled InGaAs quantum dots emitting between
918-930\,nm grown by molecular beam epitaxy (MBE) embedded in a planar low-Q
distributed Bragg reflector (DBR) cavity consisting of 24 lower and 5 upper mirror pairs. The presence of
naturally occurring, micron sized photonic defects on the sample enhances the
brightness of the photon flux\,\cite{Maier2014}.
The sample is mounted inside a helium flow cryostat and kept at a constant
temperature of 5.5\,K.  
For experiments with coherent excitation light, we use a commercial continuous wave
(cw) external cavity diode laser which is focused on to the sample using a
microscope objective (NA 0.65). A low power, non-resonant
He-Ne laser ($<$\,0.1\,nW, 637\,nm) is used fill adjacent charge traps thus
effectively gating the quantum dot fluorescence\,\cite{Nguyen2012}.
Directly reflected light is suppressed with a ratio exceeding $10^6$ by a
combination of polarization and spatial filtering prior to detection, while photons scattered by the QD are
detected by a single photon counting module (SPCM) (cf. Fig.\,1\,(b)).

For simulating the coherent excitation experiments, we follow the
semi-classical Bloch equation approach\,\cite{Allen1975}. This is well
justified, as for moderate laser powers of a few hundred nW, the average photon
number of the excitation $\braket{\hat n}$ during the lifetime of the emitter
are large and the relative photon number fluctuations $\delta n= \Delta
n/\braket{\hat n}$ can be neglected. In the present case with an excitation
power on the order of 100nW and a radiative lifetime of about 1 ns we estimate
$\braket{\hat n}$=460 and $\delta n=4.7*10^{-2}$.
Taking this estimate into account, we consider the Rabi frequency $\Omega \sim
\sqrt{\braket{\hat n}}$ fixed, i.e. not subject to quantum fluctuations. 
In this regime the resonance fluorescence intensity is directly
proportional to the average exciton population  
$\braket{\rho_{\textrm{X}}(t)}_{\textrm{pd}}$. 

\begin{figure}[htbp]  
\center
\includegraphics[width=0.5\textwidth]{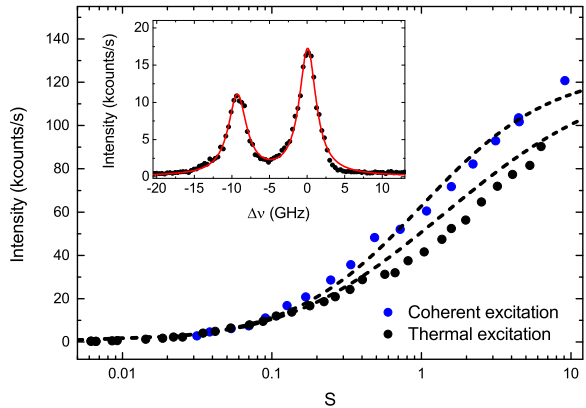} 
\caption{(Color online) Blue (Black) dots: Saturation behaviour of
TLS under coherent (chaotic) excitation. The dashed lines represent simulations
of the respective experimental conditions. The inset shows the laser scan
across resonance at an intensity of S=0.1. Two excitonic transitions are visible
with a finestructure splitting of 9.1\,GHz (37.6\,$\mu$eV). The absolute energy
at $\Delta \nu=0$ is 1.34678\,eV(920.6\,nm). The excitation power is
rescaled in units of the dimensionless saturation parameter
$S=I/I_{\textrm{sat}}=\Omega^2 T_1 T_2$ where the saturation intensity
$I_{\textrm{sat}}$ is extracted from a fit of the coherent data to Eq.\,(5).}
\label{fig:saett} 
\end{figure}

In the steady state one finds
\begin{eqnarray}
 \braket{\rho_{\textrm{X}}}_{\textrm{pd}}&=&\frac{1}{2}\frac{\Omega^{2}
T_{1}/T_{2}}{\Delta\omega^{2}+1/T_{2}^{2} +\Omega^{2}
T_{1}/T_{2}},\label{eq:saturation_curve}
\end{eqnarray}
with $\Delta\omega$ being the laser detuning with respect to exact resonance,
$T_{1}$ the exciton lifetime in the QD and $T_{2}$ the
coherence time of the exciton\,\cite{Flagg2009}.

For chaotic light with a correlation time much longer than the coherence time of
the TLS ($T_2 \leq 1$\,ns), the TLS's response can be calculated by
averaging the excited state population
$\braket{\rho_{\textrm{X}}(t)}_{\textrm{pd}}$ over the photon number
distribution given in Eq.
\ref{eq:chaotic_distribution}:
\begin{eqnarray}
\braket{\rho_{\textrm{X}}}_{\textrm{ch}}&=&\sum_{n}p_{\textrm{ch}}(n)\braket{\rho_{\textrm{X}}(n)}_{\textrm{pd}}\nonumber\\
&\approx&\int_{0}^{\infty}\mathrm d \Omega^{2}
\frac{\braket{\rho_{\textrm{X}}}_{\textrm{pd}}}{\bar{\Omega}^{2}} \exp \left(-
\Omega^{2}/\bar{\Omega}^{2}\right),
\label{eq:chaotic_exciton} \end{eqnarray}
where  $\bar{\Omega}$ is the Rabi frequency corresponding to a coherent light field with the
same average intensity and the last line holds for large average photon
numbers $\braket{\hat n}$\,\cite{Loudon2001}.
The integral in Eq.\,\ref{eq:chaotic_exciton} can
be solved analytically\,\cite{Georges1979} and it turns out that excitation of a
TLS with chaotic light is always less effective than excitation with coherent light.
This can be intuitively understood, as only one photon is absorbed to generate
an exciton and the remaining bunched photons cannot be absorbed by the TLS. The
theoretical prediction of Eq.\,\ref{eq:chaotic_exciton} is in good agreement
with the measurement shown in Fig.\,2.

\begin{figure}[htbp] 
\center
\includegraphics[width=0.5\textwidth]{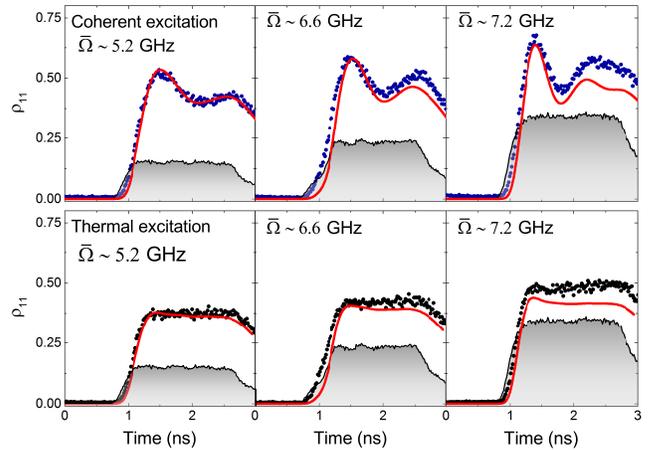} 
\caption{(Color online) Time trace (dots) of light scattered by a single QD upon
resonant excitation by 2\,ns square pulses (rescaled: filled curves). Upper
panel:
 Excitation by coherent light shows Rabi oscillations of the exciton for three
 different average Rabi frequencies $\bar{\Omega}~$(approximately 5.2, 6.6 and
 7.2\,GHz, respectively). Lower panel:
Excitation by chaotic light by pulses of the same intensity creates no
oscillations.
The solid line represent simulations based on the optical Bloch equations as
described in the main text.}
\label{fig:Rabi}
\end{figure}

In the transient regime, the well-known Rabi oscillations are the most prominent
feature of two level systems interacting with a coherent field.
While quantum fluctuations of a coherent field can in principle lead to marked deviations from the
 classical light field, 
e.g. the collapse and subsequent revival of Rabi
oscillations\,\cite{Rempe1987,Mukamel2015a}, their influence on our experiments
is negligible as discussed above. 
For chaotic light, this regime has been studied theoretically and it has
been predicted that Rabi oscillations should be suppressed by the fluctuations
present in chaotic fields\,\cite{Knight1982}.
To experimentally verify these predictions, we use an electro-optical modulator
(EOM) to temporally shape the emission of the cw light source into square pulses
with a length of 2\,ns and a repetition rate of 10\,MHz.
The arrival times of photons scattered by the QD are recorded and histogrammed
over an integration time of a few minutes.
The measurements under coherent resonant excitation of the TLS are depicted in
the upper panel in Fig.\,3. They show clear Rabi oscillations being damped by
radiative and pure dephasing present in the solid state system\,\cite{Ramsay2010,Kuhlmann2015}. 
This result is in excellent agreement with the numerical solutions of the
semi-classical Bloch equations, where $T_2=(325\pm5)$\,ps and
$T_1=(641\pm 62)$\,ps were determined from independent linewidth and
$g^{(2)}(\tau)$ measurements, respectively.
In stark contrast, the time resolved fluorescence signal upon chaotic
excitation bears no signatures of coherence generated in the TLS. This is a direct
 consequence of the pronounced intensity fluctuations of the chaotic field.

\begin{figure}[htbp]
\center 
\includegraphics[width=0.5\textwidth]{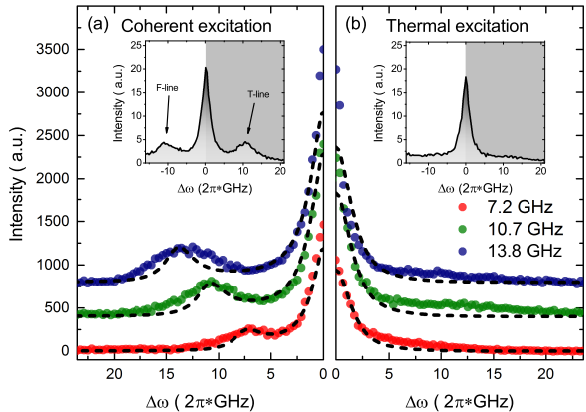}  
\caption{(Color online) Emission spectrum of the strongly driven QD for three
different average Rabi frequencies. The panel (a) on the left displays the
results obtained for coherent excitation. The T-line of the Mollow triplett is clearly visible.
Inset: Complete emission spectrum highlighting displayed part of Mollow triplet.
For chaotic excitation, shown in panel (b), no sidepeaks are discernible.
The dashed lines represent simulations corresponding to the experimental
conditions.}
\label{fig:Mollow}
\end{figure}  
Besides Rabi oscillations, the iconic Mollow triplet is a further
hallmark of resonance fluorescence using coherent excitation.
 It consists of one central peak
(at frequency $\nu _0$) and two symmetrically shifted satellite peaks (at $\nu
_0 \pm \Omega $). It is a consequence of the interaction of a two level system
with an intense coherent light field and can be handily interpreted in
the framework of dressed states as was first proposed by Cohen-Tannoudji et al.\,\cite{Cohen-Tannoudji1977}.
However, this picture only holds in the the case of $\braket{\hat n} \gg
\braket{\Delta \hat{n}} \gg 1$, which is true for a coherent state but not
for chaotic light where $\braket{\hat n} = \braket{\Delta \hat{n}}$. 
Thus, for the chaotic case, it has been predicted that the two satellite peaks
should disappear\,\cite{Avan1977,Georges1979}.

To experimentally observe the satellite peaks which 
are purely part of the incoherently scattered fraction of the total fluorescence
at high excitation power we use a scanning Fabry-Perot resonator with a free
spectra range of 26.4\,GHz (109.4\,$\mu eV$) and a resolution of 175.4\,MHz
(725.4\,neV).
Plotted in Fig.\,4\,(a) is the  right wing of the Mollow triplet (T-line) under
strict resonant excitation for three different average Rabi frequencies.
The satellite peak is clearly visible under coherent excitation.
The experimental data is in good agreement with the predicted power spectrum
including pure dephasing \cite{Matthiesen2012}.
Excitation induced dephasing which leads to a broadening of the Mollow
sidepeaks at high excitation power \cite{Ulrich2011}, is not included in the
theory and therefore likely to cause the deviations between theory and experiment at higher
Rabi frequencies.

In contrast, under chaotic excitation [Fig.\,4\,(b)] the Mollow triplet cannot
be observed under otherwise identical excitation conditions.
This is again a direct consequence of the large intensity fluctuations present
in the chaotic light field and can be quantitatively explained
by averaging the power dependent spectra over the  photon number distribution
given in Eq.\,(3). %
In the experiments, this is achieved by integrating over times at least 5 orders
of magnitude longer than the correlation time of our chaotic light source. This
ensures that a thermal photon number distribution is sampled over the
course of each integration interval meaning also that we average over the entire
range of Rabi frequencies present under chaotic excitation.

While the previous experiments have shown that the interaction of a TLS with a
light field differs significantly depending on the photon statistics inherent in
the exciting light field, it is also very interesting to explore its influence
on the emission statistics of the TLS.
\begin{figure}[htbp]
\center
\includegraphics[width=0.45\textwidth]{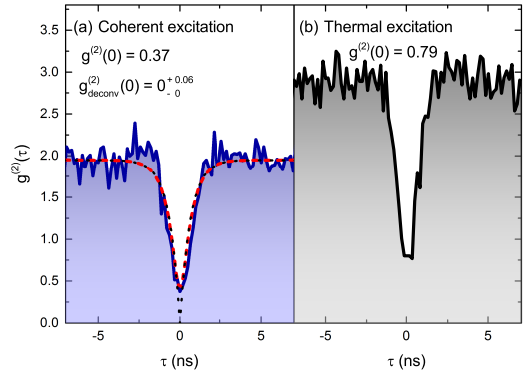} 
\caption{(Color online) $g^{(2)}(\tau)$-measurement of the emitted radiation.
The excitonic transition is driven at a Rabi frequency of
$\Omega\approx1.7$\,GHz (0.6\,S). (a) Measurement using the laser as light
source (solid line). The dashed line shows a convolution of the
solution for a two-level system with the detector response and provides very
good agreement with the experimental data. (b) Measurement using the Martienssen
lamp as light source. 
In both measurements pronounced anti-bunching is visible at $\tau=0$.}
\label{fig:g2}
\end{figure}

For this purpose, a Hanbury-Brown
and Twiss setup consisting of a fiber-based 50:50 beam splitter and two
SPCMs (timing resolution 351\,ps) is used to measure the second-order
autocorrelation function.
In Fig.\,5\,(a), the measured $g^{(2)}(\tau)$, normalized to the average
count rate during the experiment under coherent excitation is
shown. Superimposed onto the antibunching dip at $\tau=0$, as is expected for a single TLS,
 pronounced bunching is observed for larger $\tau$.
This is typical for resonance fluorescence experiments on QDs and is attributed
to blinking of the QD\,\cite{Ulhaq2012}.   
In Fig.\,5\,(b) the same measurement is shown for chaotic excitation. Here, the
antibunching is visible with an increased bunching compared to the coherent case.
Thus, as intuitively expected, the
non-classical nature of the emitted radiation is preserved irrespective of the
photon statistics of the exciting light field. 
While for an ideal TLS a quasi-stationary value of 2 is expected for $
T_1<\tau<\tau_{\textrm{corr}}$ under chaotic excitation \cite{Schubert1979}, we
observe an increased bunching ($g^{(2)}(\tau)=3$). This is probably
caused by the blinking behaviour of the QD already visible under coherent excitation.
Interestingly, the different bunching behaviour under thermal and
coherent excitation indicates that the photon statistics of the excitation
influences the carrier distribution and occupation dynamics of QDs which could
be a topic of further investigations beyond the scope of the present work.
In this regard, it is also noteworthy that without careful renormalisation of the autocorrelation data no direct
 difference between the two types of excitation would be observable.
 
In conclusion, our experiments show that the response of a quantum mechanical two level system is very sensitive
to the photon statistics of the exciting light field. While differences are already visible
in the saturation behavior of the TLS, the more striking differences occur in
the transient regime, where Rabi oscillations are suppressed under chaotic
excitation. 
Furthermore, the emission spectrum under strong excitation depends dramatically
on the higher order correlation functions of the exciting light field. Thus, the
iconic Mollow triplet disappears under chaotic excitation.
With its nonlinear nature, the fermionic TLS is an ideal probe for the fluctuations present in the light field.
Future experiments will be directed towards exploring the regime of short correlation times in the excitation field,
being on the order of or even shorter than the coherence time of the TLS. Also,
extending photon-statistics excitation spectroscopy to non-classical light
sources will be highly interesting.

We thank M. A\ss mann and A. Carmele for stimulating discussions. 
The research leading to these results has received funding from from the
European Research Council under the European Union's Seventh Framework ERC Grant Agreement
No. 615613 and from the German Research Foundation via the project RE2974/5-1.

\section*{\underline{Supplemental material:} Photon-statistics excitation
spectroscopy of a single two-level system}

\renewcommand{\thefigure}{S\arabic{figure}}

\setcounter{figure}{0}
\section{Emission spectrum}
\begin{figure}[htbp] 
\center
\includegraphics[width=0.5\textwidth]{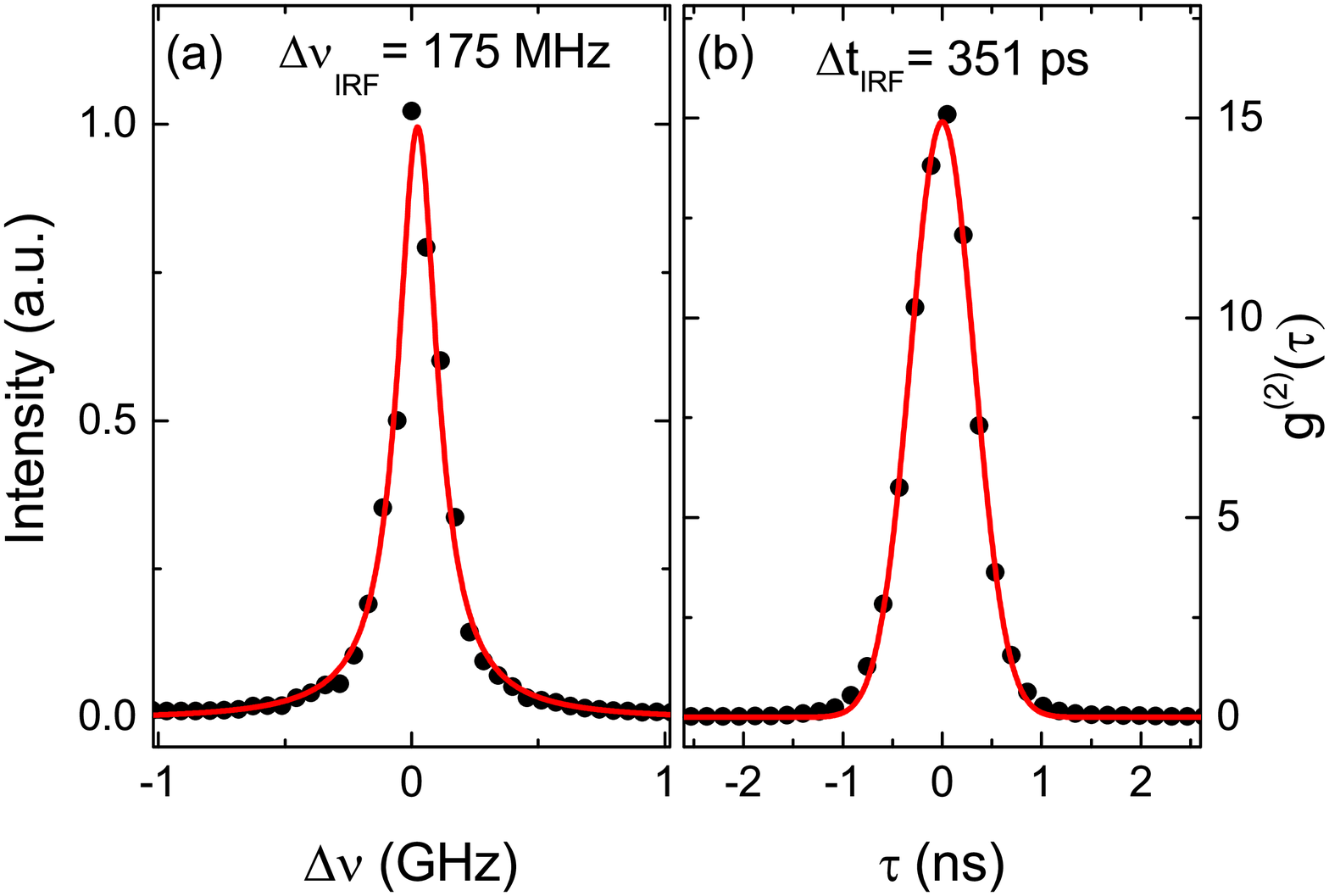} 
\caption{(Color online) Instrument response functions for FPI (a) and
intensity autocorrelation (b) measurements. (a) Response of the FPI to the cw-
diode laser ($\Delta \omega \approx 100$\,kHz) is shown fitted with a Lorentzian
yielding a FWHM of 175\,MHz.
(b) Response of the intensity autocorrelation measurement (black dots) yielding
a $\Delta \tau_{IRF}= 351$\,ps as extracted from a Gaussian fit to the data. }
\label{figs1}
\end{figure}

The emission spectra shown in the main text were recorded using a
custom-built closed-loop scanning Fabry-Perot cavity (FPI) with a spectral
resolution of 175\,MHz. The instrument response function (IRF)
of the FPI is displayed in Fig.\,\ref{figs1}.
% The analytical solution of the
%emission spectrum of a TLS under coherent excitation is given by
For the analysis of the experimental
results, a convolution of the analytical solution for the Mollow triplet with
the IRF was fitted to the data obtained under coherent excitation.
% \begin{widetext}
% \begin{equation}
% S(\Delta \nu) \propto \frac{1}{2}\frac{\textrm{T}_2^{-1}}{\Delta \nu ^2
% +T_2^{-2}}+ \frac{1}{2}\frac{1}{(\textrm{T}_1
% T_2)^{-1}+\Omega^2}(\frac{A\frac{\eta}{2}-(8\mu)^{-1}(\Delta \nu-\mu)B}{(\Delta
% \nu - \mu)^2 + \eta^2}+\frac{A\frac{\eta}{2}+(8\mu)^{-1}(\Delta
% \nu+\mu)B}{(\Delta \nu + \mu)^2 + \eta^2}) + \frac{1}{T_1 T_2^{-1}+ \Omega ^2
% T_{1}^2} \delta (\Delta \nu) ,\label{eq:Mollow_triplet}
% \end{equation}. 
% \end{widetext}
% Here, $A =$,$B=$, $\mu=$, $\eta=$ . $T_1$ and $T_2$ are the longitudinal and
% transverse relaxation times, respectively. 

%\section{Experimental setup}
% The data of the emission spectra in the main text were recorded using a FPI with
% a spectral resolution of 175 MHz. Be
% The FPI and and HBT setup used for spectral and autocorrelation measurements,
% respectively, both have  The instrument response function (IRF)  
\section{Rabi oscillations}
The square pulses used for this experiment were generated using a
fiber-based EOM which was voltage modulated using an
arbitrary wave form generator (AWG) with a bandwidth of 5 GHz. In order to
capture the entire temporal evolution of the TLS we chose a pulse length of
2\,ns which is substantially longer than the lifetime ($T_1=641$\,ps) of the TLS.
The photons scattered by the QD were recorded using a Silicon-based avalanche photo diode and time-correlated single photon counting (TCSPC) electronics affording a temporal resolution of 40\,ps.
\begin{figure}[htbp] 
\center
\includegraphics[width=0.5\textwidth]{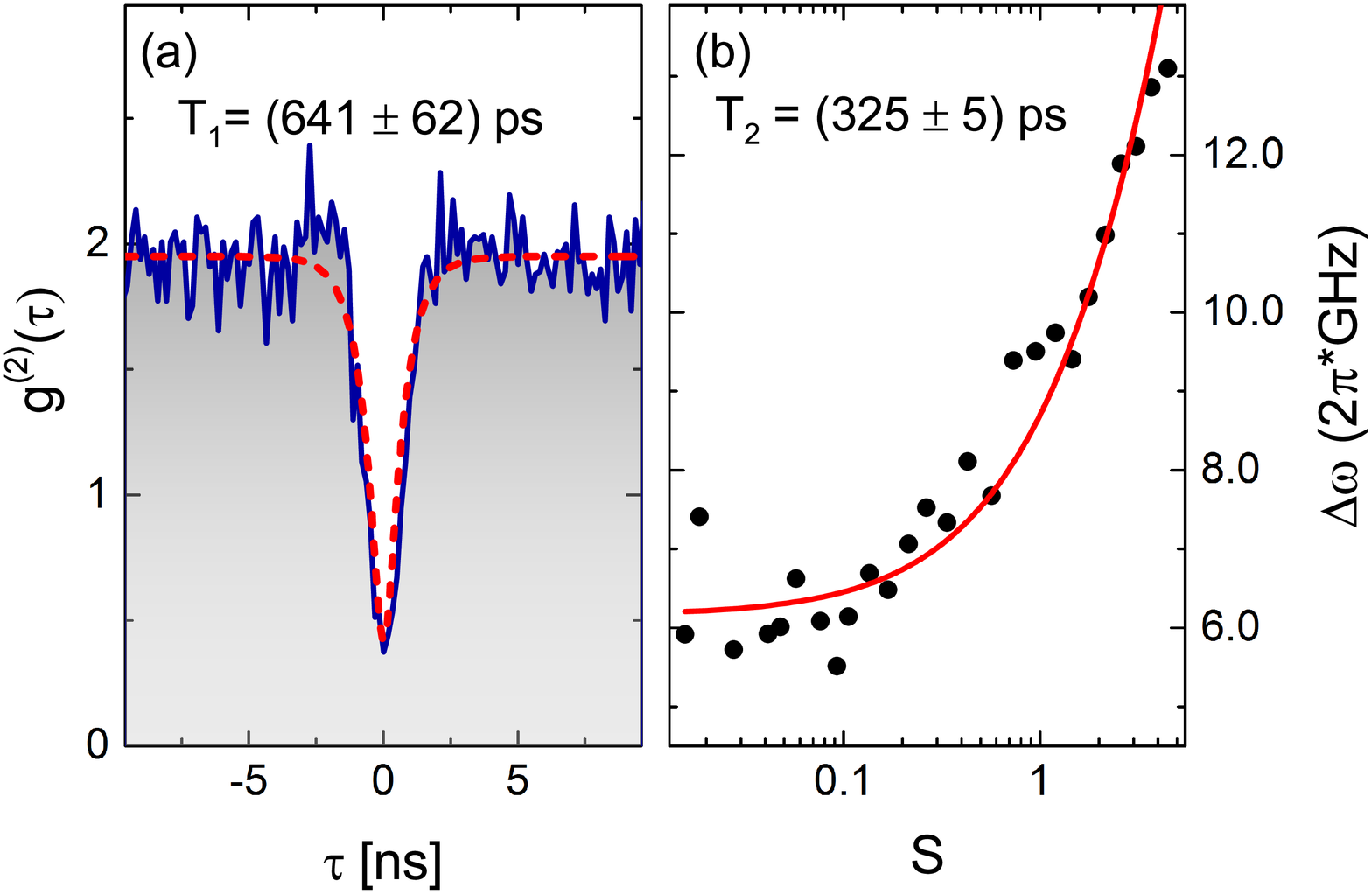} 
\caption{(Color online) Measurements performed for determining $T_{1}$ and
$T_{2}$. (a) Intensity autocorrelation measurement of photons emitted by the QD.
A fit of Eq. \ref{eqn:g2} convolved with the IRF to the data yields
$T_1=641$\,ps.
(b) Dependence of the linewidth of the TLS on the excitation power (black dots). The solid line represents a fit of Eq.
\ref{eqn:pdlw} to the data giving $T_2=325$\,ps. }
\label{figs2}
\end{figure}
The AWG supplied a trigger signal to the TCSPC electronics while the measurement
was stopped by the detection of a photon. The measurement was simulated using the standard
semi-classical Bloch equations where $\rho_{11}=1-\rho_{00}$
($\rho_{01}=\rho_{10}^{*}$) represent the diagonal (off-diagonal) elements of
the density matrix of the TLS:
\begin{eqnarray}
\dot{\rho}_{11}(t)=i \frac{\Omega_R(t)}{t}
(\rho_{10}-\rho_{01})-\frac{1}{T_1}\rho_{11}(t)\label{eq:BGen1}\\
\dot{\rho}_{01}(t)=-(i(\Delta\omega)+\frac{1}{T2})\rho_{01}(t)-i\frac{\Omega_R(t)(\rho_{11}-\rho_{00})}{2})
\label{eq:BGen2}
\end{eqnarray}
Here, $\Delta \omega =\omega_l-\omega_0$ denotes the detuning between laser and
exciton transition which is zero for all experiments conducted in this study and
$\Omega_R(t)$ the time dependent Rabi frequency. The experimentally obtained
data (APD countrates) were converted into a normalised occupation by fitting the
data to the solution of equation \ref{eq:BGen1} with only free parameter being
the conversion factor. The system properties, i.e.
the dephasing time $T_2$ and population decay time $T_1$, were determined
independently from the experiments shown in Fig.
\ref{figs2} as follows. A $g^{(2)}(\tau)$-measurement of the photons emitted
by the TLS driven at a low Rabi frequency allows to extract the population decay
time $T_1$. In this limiting case, the second-order autocorrelation function is
given by
\begin{equation}
g^{(2)}(\tau)=1-\textrm{exp}(-\frac{|\tau|}{T_1})~. \
\label{eqn:g2}
\end{equation}
Fitting this formula convolved with the IRF to the data presented in
Fig.\ref{figs2} (a) yields $T_1=(641 \pm 64$)\,ps.

Power dependent linewidth measurements of the exciton which are shown in Fig.\,
\ref{figs2}\,(b) yield the $T_2$ time. The power dependence of the linewidth of
a TLS is given by \begin{equation}\Delta \omega = \frac{2}{T_2}\sqrt{1+\Omega ^2 T_1
T_2} \label{eqn:pdlw}~. \end{equation}
For low intensities, i.e. $\Omega^2 \rightarrow 0$, this expression tends
towards $\Delta \omega = \frac{2}{T_2}.$
%  \[\Omega_R(t)=\sum_{i=1}^{5} 
% a_{i}e^{\frac{-(t-t_i)^2}{4 c^2}}\] where $a_i$ and $t_i$ are determined by
% fitting the experimentally recorded excitation pulses with 5 Gaussians.
\section{$g^{(2)}(\tau)$-measurement}  
The autocorrelation measurements were performed using a Hanbury-Brown and Twiss
setup with an overall timing resolution of 351\,ps (cf. Fig.\,\ref{figs1}\,(b)).
\begin{figure}[htbp]
\center
\includegraphics[width=0.5\textwidth]{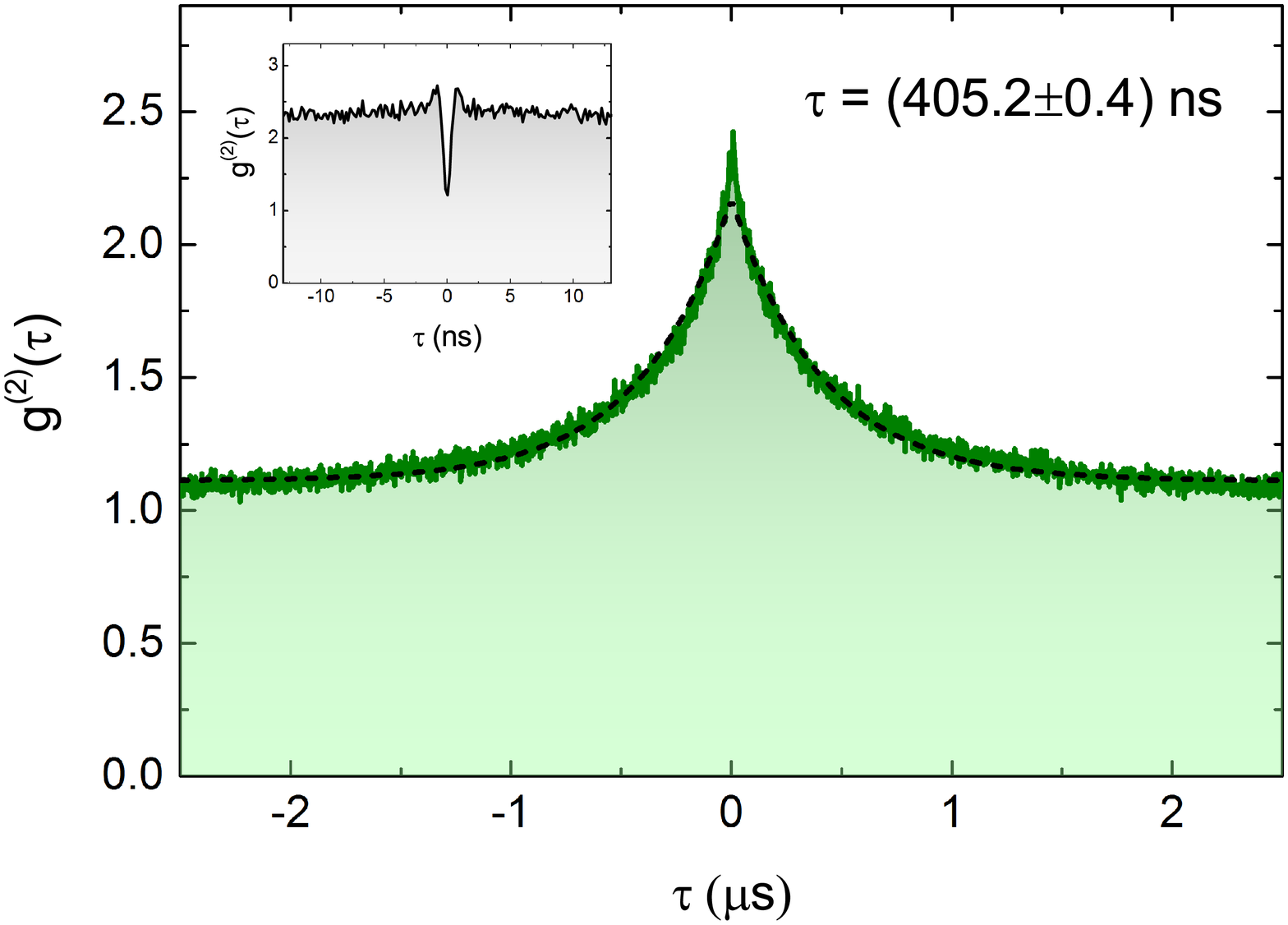} 
\caption{(Color online) Intensity autocorrelation measurement showing pronounced
bunching on long timescales under coherent excitation with a Rabi frequency of
7.1\,GHz (S=10.6).
A bidirectional exponential is fit to the data (dashed line) giving a time constant $\tau=405$\,ns. Inset:
Antibunching is observed on a ns timescale where the onset of Rabi oscillations
are visible.
}
\label{figs3}
\end{figure}

The recorded raw data $c(\tau)$ was converted to the normalised $g^{(2)}(\tau)$
according to the following equation $C_{N}(\tau)=\frac{c(\tau)}{N_1 N_2 w T}$
where $N_{i}$ are the counts detected by APD $i=\{1,2\}$, w is the width of the
bins and T is the integration time. The resonance fluorescence of the 4 quantum
dots studied in this sample all exhibited bunching on the time scale of several
100\,ns as is illustrated by the measurement in Fig.\,\ref{figs3}. Furthermore,
all of them showed an increased bunching under thermal excitation as
discussed in the main text.

%\section{The Martiennsen lamp}

\bibliography{QTS_TLS.bib}

%merlin.mbs apsrev4-1.bst 2010-07-25 4.21a (PWD, AO, DPC) hacked
%Control: key (0)
%Control: author (8) initials jnrlst
%Control: editor formatted (1) identically to author
%Control: production of article title (-1) disabled
%Control: page (0) single
%Control: year (1) truncated
%Control: production of eprint (0) enabled
\begin{thebibliography}{31}%
\makeatletter
\providecommand \@ifxundefined [1]{%
 \@ifx{#1\undefined}
}%
\providecommand \@ifnum [1]{%
 \ifnum #1\expandafter \@firstoftwo
 \else \expandafter \@secondoftwo
 \fi
}%
\providecommand \@ifx [1]{%
 \ifx #1\expandafter \@firstoftwo
 \else \expandafter \@secondoftwo
 \fi
}%
\providecommand \natexlab [1]{#1}%
\providecommand \enquote  [1]{``#1''}%
\providecommand \bibnamefont  [1]{#1}%
\providecommand \bibfnamefont [1]{#1}%
\providecommand \citenamefont [1]{#1}%
\providecommand \href@noop [0]{\@secondoftwo}%
\providecommand \href [0]{\begingroup \@sanitize@url \@href}%
\providecommand \@href[1]{\@@startlink{#1}\@@href}%
\providecommand \@@href[1]{\endgroup#1\@@endlink}%
\providecommand \@sanitize@url [0]{\catcode `\\12\catcode `\$12\catcode
  `\&12\catcode `\#12\catcode `\^12\catcode `\_12\catcode `\%12\relax}%
\providecommand \@@startlink[1]{}%
\providecommand \@@endlink[0]{}%
\providecommand \url  [0]{\begingroup\@sanitize@url \@url }%
\providecommand \@url [1]{\endgroup\@href {#1}{\urlprefix }}%
\providecommand \urlprefix  [0]{URL }%
\providecommand \Eprint [0]{\href }%
\providecommand \doibase [0]{http://dx.doi.org/}%
\providecommand \selectlanguage [0]{\@gobble}%
\providecommand \bibinfo  [0]{\@secondoftwo}%
\providecommand \bibfield  [0]{\@secondoftwo}%
\providecommand \translation [1]{[#1]}%
\providecommand \BibitemOpen [0]{}%
\providecommand \bibitemStop [0]{}%
\providecommand \bibitemNoStop [0]{.\EOS\space}%
\providecommand \EOS [0]{\spacefactor3000\relax}%
\providecommand \BibitemShut  [1]{\csname bibitem#1\endcsname}%
\let\auto@bib@innerbib\@empty
%</preamble>
\bibitem [{\citenamefont {Devoret}\ and\ \citenamefont
  {Schoelkopf}(2013)}]{Devoret2013a}%
  \BibitemOpen
  \bibfield  {author} {\bibinfo {author} {\bibfnamefont {M.~H.}\ \bibnamefont
  {Devoret}}\ and\ \bibinfo {author} {\bibfnamefont {R.~J.}\ \bibnamefont
  {Schoelkopf}},\ }\href {\doibase 10.1126/science.1231930} {\bibfield
  {journal} {\bibinfo  {journal} {Science}\ }\textbf {\bibinfo {volume}
  {339}},\ \bibinfo {pages} {1169} (\bibinfo {year} {2013})}\BibitemShut
  {NoStop}%
\bibitem [{\citenamefont {Wrachtrup}\ and\ \citenamefont
  {Jelezko}(2006)}]{0953-8984-18-21-S08}%
  \BibitemOpen
  \bibfield  {author} {\bibinfo {author} {\bibfnamefont {J.}~\bibnamefont
  {Wrachtrup}}\ and\ \bibinfo {author} {\bibfnamefont {F.}~\bibnamefont
  {Jelezko}},\ }\href {\doibase 10.1088/0953-8984/18/21/S08} {\bibfield
  {journal} {\bibinfo  {journal} {J. Phys. Condens. Matter}\ }\textbf {\bibinfo
  {volume} {18}},\ \bibinfo {pages} {S807} (\bibinfo {year}
  {2006})}\BibitemShut {NoStop}%
\bibitem [{\citenamefont {Nielsen}\ and\ \citenamefont
  {Chuang}(2010)}]{Nielsen2010}%
  \BibitemOpen
  \bibfield  {author} {\bibinfo {author} {\bibfnamefont {M.}~\bibnamefont
  {Nielsen}}\ and\ \bibinfo {author} {\bibfnamefont {I.}~\bibnamefont
  {Chuang}},\ }\href
  {http://books.google.com/books?hl=en&lr=&id=-s4DEy7o-a0C&oi=fnd&pg=PR17&dq=Quantum+Computation+and+Quantum+Information&ots=NF7EgrsAYq&sig=dr818XRE_rrmrQ-oxFVSEdw3etI}
  {\emph {\bibinfo {title} {{Quantum Computation and Quantum Information}}}}\
  (\bibinfo  {publisher} {Cambridge University Press},\ \bibinfo {address}
  {Camebridge},\ \bibinfo {year} {2010})\BibitemShut {NoStop}%
\bibitem [{\citenamefont {Leibfried}\ \emph {et~al.}(2003)\citenamefont
  {Leibfried}, \citenamefont {Blatt}, \citenamefont {Monroe},\ and\
  \citenamefont {Wineland}}]{Leibfried2003}%
  \BibitemOpen
  \bibfield  {author} {\bibinfo {author} {\bibfnamefont {D.}~\bibnamefont
  {Leibfried}}, \bibinfo {author} {\bibfnamefont {R.}~\bibnamefont {Blatt}},
  \bibinfo {author} {\bibfnamefont {C.}~\bibnamefont {Monroe}}, \ and\ \bibinfo
  {author} {\bibfnamefont {D.}~\bibnamefont {Wineland}},\ }\href {\doibase
  10.1103/RevModPhys.75.281} {\bibfield  {journal} {\bibinfo  {journal} {Rev.
  Mod. Phys.}\ }\textbf {\bibinfo {volume} {75}},\ \bibinfo {pages} {281}
  (\bibinfo {year} {2003})}\BibitemShut {NoStop}%
\bibitem [{\citenamefont {Meyer}\ \emph {et~al.}(2015)\citenamefont {Meyer},
  \citenamefont {Stockill}, \citenamefont {Steiner}, \citenamefont {{Le Gall}},
  \citenamefont {Matthiesen}, \citenamefont {Clarke}, \citenamefont {Ludwig},
  \citenamefont {Reichel}, \citenamefont {Atat{\"{u}}re},\ and\ \citenamefont
  {K{\"{o}}hl}}]{Meyer2015}%
  \BibitemOpen
  \bibfield  {author} {\bibinfo {author} {\bibfnamefont {H.~M.}\ \bibnamefont
  {Meyer}}, \bibinfo {author} {\bibfnamefont {R.}~\bibnamefont {Stockill}},
  \bibinfo {author} {\bibfnamefont {M.}~\bibnamefont {Steiner}}, \bibinfo
  {author} {\bibfnamefont {C.}~\bibnamefont {{Le Gall}}}, \bibinfo {author}
  {\bibfnamefont {C.}~\bibnamefont {Matthiesen}}, \bibinfo {author}
  {\bibfnamefont {E.}~\bibnamefont {Clarke}}, \bibinfo {author} {\bibfnamefont
  {A.}~\bibnamefont {Ludwig}}, \bibinfo {author} {\bibfnamefont
  {J.}~\bibnamefont {Reichel}}, \bibinfo {author} {\bibfnamefont
  {M.}~\bibnamefont {Atat{\"{u}}re}}, \ and\ \bibinfo {author} {\bibfnamefont
  {M.}~\bibnamefont {K{\"{o}}hl}},\ }\href {\doibase
  10.1103/PhysRevLett.114.123001} {\bibfield  {journal} {\bibinfo  {journal}
  {Phys. Rev. Lett.}\ }\textbf {\bibinfo {volume} {114}},\ \bibinfo {pages}
  {123001} (\bibinfo {year} {2015})}\BibitemShut {NoStop}%
\bibitem [{\citenamefont {Wolters}\ \emph {et~al.}(2013)\citenamefont
  {Wolters}, \citenamefont {Strau{\ss}}, \citenamefont {Schoenfeld},\ and\
  \citenamefont {Benson}}]{Wolters2013a}%
  \BibitemOpen
  \bibfield  {author} {\bibinfo {author} {\bibfnamefont {J.}~\bibnamefont
  {Wolters}}, \bibinfo {author} {\bibfnamefont {M.}~\bibnamefont {Strau{\ss}}},
  \bibinfo {author} {\bibfnamefont {R.~S.}\ \bibnamefont {Schoenfeld}}, \ and\
  \bibinfo {author} {\bibfnamefont {O.}~\bibnamefont {Benson}},\ }\href
  {\doibase 10.1103/PhysRevA.88.020101} {\bibfield  {journal} {\bibinfo
  {journal} {Phys. Rev. A}\ }\textbf {\bibinfo {volume} {88}},\ \bibinfo
  {pages} {020101} (\bibinfo {year} {2013})}\BibitemShut {NoStop}%
\bibitem [{\citenamefont {Georgiades}\ \emph {et~al.}(1995)\citenamefont
  {Georgiades}, \citenamefont {Polzik}, \citenamefont {Edamatsu}, \citenamefont
  {Kimble},\ and\ \citenamefont {Parkins}}]{Georgiades1995}%
  \BibitemOpen
  \bibfield  {author} {\bibinfo {author} {\bibfnamefont {N.~P.}\ \bibnamefont
  {Georgiades}}, \bibinfo {author} {\bibfnamefont {E.~S.}\ \bibnamefont
  {Polzik}}, \bibinfo {author} {\bibfnamefont {K.}~\bibnamefont {Edamatsu}},
  \bibinfo {author} {\bibfnamefont {H.~J.}\ \bibnamefont {Kimble}}, \ and\
  \bibinfo {author} {\bibfnamefont {A.~S.}\ \bibnamefont {Parkins}},\ }\href
  {\doibase 10.1103/PhysRevLett.75.3426} {\bibfield  {journal} {\bibinfo
  {journal} {Phys. Rev. Lett.}\ }\textbf {\bibinfo {volume} {75}},\ \bibinfo
  {pages} {3426} (\bibinfo {year} {1995})}\BibitemShut {NoStop}%
\bibitem [{\citenamefont {Murch}\ \emph {et~al.}(2013)\citenamefont {Murch},
  \citenamefont {Weber}, \citenamefont {Beck}, \citenamefont {Ginossar},\ and\
  \citenamefont {Siddiqi}}]{Murch2013}%
  \BibitemOpen
  \bibfield  {author} {\bibinfo {author} {\bibfnamefont {K.~W.}\ \bibnamefont
  {Murch}}, \bibinfo {author} {\bibfnamefont {S.~J.}\ \bibnamefont {Weber}},
  \bibinfo {author} {\bibfnamefont {K.~M.}\ \bibnamefont {Beck}}, \bibinfo
  {author} {\bibfnamefont {E.}~\bibnamefont {Ginossar}}, \ and\ \bibinfo
  {author} {\bibfnamefont {I.}~\bibnamefont {Siddiqi}},\ }\href {\doibase
  10.1038/nature12264} {\bibfield  {journal} {\bibinfo  {journal} {Nature}\
  }\textbf {\bibinfo {volume} {499}},\ \bibinfo {pages} {62} (\bibinfo {year}
  {2013})}\BibitemShut {NoStop}%
\bibitem [{\citenamefont {Carmele}\ \emph {et~al.}(2009)\citenamefont
  {Carmele}, \citenamefont {Knorr},\ and\ \citenamefont
  {Richter}}]{Carmele2009}%
  \BibitemOpen
  \bibfield  {author} {\bibinfo {author} {\bibfnamefont {A.}~\bibnamefont
  {Carmele}}, \bibinfo {author} {\bibfnamefont {A.}~\bibnamefont {Knorr}}, \
  and\ \bibinfo {author} {\bibfnamefont {M.}~\bibnamefont {Richter}},\ }\href
  {\doibase 10.1103/PhysRevB.79.035316} {\bibfield  {journal} {\bibinfo
  {journal} {Phys. Rev. B}\ }\textbf {\bibinfo {volume} {79}},\ \bibinfo
  {pages} {035316} (\bibinfo {year} {2009})}\BibitemShut {NoStop}%
\bibitem [{\citenamefont {Kazimierczuk}\ \emph {et~al.}(2015)\citenamefont
  {Kazimierczuk}, \citenamefont {Schmutzler}, \citenamefont {A{\ss}mann},
  \citenamefont {Schneider}, \citenamefont {Kamp}, \citenamefont
  {H{\"{o}}fling},\ and\ \citenamefont {Bayer}}]{Kazimierczuk2015}%
  \BibitemOpen
  \bibfield  {author} {\bibinfo {author} {\bibfnamefont {T.}~\bibnamefont
  {Kazimierczuk}}, \bibinfo {author} {\bibfnamefont {J.}~\bibnamefont
  {Schmutzler}}, \bibinfo {author} {\bibfnamefont {M.}~\bibnamefont
  {A{\ss}mann}}, \bibinfo {author} {\bibfnamefont {C.}~\bibnamefont
  {Schneider}}, \bibinfo {author} {\bibfnamefont {M.}~\bibnamefont {Kamp}},
  \bibinfo {author} {\bibfnamefont {S.}~\bibnamefont {H{\"{o}}fling}}, \ and\
  \bibinfo {author} {\bibfnamefont {M.}~\bibnamefont {Bayer}},\ }\href
  {\doibase 10.1103/PhysRevLett.115.027401} {\bibfield  {journal} {\bibinfo
  {journal} {Phys. Rev. Lett.}\ }\textbf {\bibinfo {volume} {115}},\ \bibinfo
  {pages} {027401} (\bibinfo {year} {2015})}\BibitemShut {NoStop}%
\bibitem [{\citenamefont {Boitier}\ \emph {et~al.}(2009)\citenamefont
  {Boitier}, \citenamefont {Godard}, \citenamefont {Rosencher},\ and\
  \citenamefont {Fabre}}]{Boitier2009}%
  \BibitemOpen
  \bibfield  {author} {\bibinfo {author} {\bibfnamefont {F.}~\bibnamefont
  {Boitier}}, \bibinfo {author} {\bibfnamefont {A.}~\bibnamefont {Godard}},
  \bibinfo {author} {\bibfnamefont {E.}~\bibnamefont {Rosencher}}, \ and\
  \bibinfo {author} {\bibfnamefont {C.}~\bibnamefont {Fabre}},\ }\href
  {\doibase 10.1038/nphys1218} {\bibfield  {journal} {\bibinfo  {journal} {Nat.
  Phys.}\ }\textbf {\bibinfo {volume} {5}},\ \bibinfo {pages} {267} (\bibinfo
  {year} {2009})}\BibitemShut {NoStop}%
\bibitem [{\citenamefont {Loudon}(2001)}]{Loudon2001}%
  \BibitemOpen
  \bibfield  {author} {\bibinfo {author} {\bibfnamefont {R.}~\bibnamefont
  {Loudon}},\ }\href {\doibase 10.1080/716099348} {\emph {\bibinfo {title}
  {{The Quantum Theory of Light}}}},\ \bibinfo {edition} {3rd}\ ed.\ (\bibinfo
  {publisher} {Oxford University Press},\ \bibinfo {address} {Oxford},\
  \bibinfo {year} {2001})\BibitemShut {NoStop}%
\bibitem [{\citenamefont {Martienssen}\ and\ \citenamefont
  {Spiller}(1964)}]{Martienssen1964}%
  \BibitemOpen
  \bibfield  {author} {\bibinfo {author} {\bibfnamefont {W.}~\bibnamefont
  {Martienssen}}\ and\ \bibinfo {author} {\bibfnamefont {E.}~\bibnamefont
  {Spiller}},\ }\href {\doibase 10.1119/1.1970023} {\bibfield  {journal}
  {\bibinfo  {journal} {Am. J. Phys.}\ }\textbf {\bibinfo {volume} {32}},\
  \bibinfo {pages} {919} (\bibinfo {year} {1964})}\BibitemShut {NoStop}%
\bibitem [{\citenamefont {Arecchi}(1965)}]{Arecchi1965}%
  \BibitemOpen
  \bibfield  {author} {\bibinfo {author} {\bibfnamefont {F.~T.}\ \bibnamefont
  {Arecchi}},\ }\href {\doibase 10.1103/PhysRevLett.15.912} {\bibfield
  {journal} {\bibinfo  {journal} {Phys. Rev. Lett.}\ }\textbf {\bibinfo
  {volume} {15}},\ \bibinfo {pages} {912} (\bibinfo {year} {1965})}\BibitemShut
  {NoStop}%
\bibitem [{\citenamefont {Martienssen}\ and\ \citenamefont
  {Spiller}(1966)}]{Martienssen1966}%
  \BibitemOpen
  \bibfield  {author} {\bibinfo {author} {\bibfnamefont {W.}~\bibnamefont
  {Martienssen}}\ and\ \bibinfo {author} {\bibfnamefont {E.}~\bibnamefont
  {Spiller}},\ }\href@noop {} {\bibfield  {journal} {\bibinfo  {journal} {Phys.
  Rev. Lett.}\ }\textbf {\bibinfo {volume} {16}},\ \bibinfo {pages} {531}
  (\bibinfo {year} {1966})}\BibitemShut {NoStop}%
\bibitem [{\citenamefont {Maier}\ \emph {et~al.}(2014)\citenamefont {Maier},
  \citenamefont {Gold}, \citenamefont {Forchel}, \citenamefont {Gregersen},
  \citenamefont {M{\o}rk}, \citenamefont {H{\"{o}}fling}, \citenamefont
  {Schneider},\ and\ \citenamefont {Kamp}}]{Maier2014}%
  \BibitemOpen
  \bibfield  {author} {\bibinfo {author} {\bibfnamefont {S.}~\bibnamefont
  {Maier}}, \bibinfo {author} {\bibfnamefont {P.}~\bibnamefont {Gold}},
  \bibinfo {author} {\bibfnamefont {A.}~\bibnamefont {Forchel}}, \bibinfo
  {author} {\bibfnamefont {N.}~\bibnamefont {Gregersen}}, \bibinfo {author}
  {\bibfnamefont {J.}~\bibnamefont {M{\o}rk}}, \bibinfo {author} {\bibfnamefont
  {S.}~\bibnamefont {H{\"{o}}fling}}, \bibinfo {author} {\bibfnamefont
  {C.}~\bibnamefont {Schneider}}, \ and\ \bibinfo {author} {\bibfnamefont
  {M.}~\bibnamefont {Kamp}},\ }\href {\doibase 10.1364/OE.22.008136} {\bibfield
   {journal} {\bibinfo  {journal} {Opt. Express}\ }\textbf {\bibinfo {volume}
  {22}},\ \bibinfo {pages} {8136} (\bibinfo {year} {2014})}\BibitemShut
  {NoStop}%
\bibitem [{\citenamefont {Nguyen}\ \emph {et~al.}(2012)\citenamefont {Nguyen},
  \citenamefont {Sallen}, \citenamefont {Voisin}, \citenamefont {Roussignol},
  \citenamefont {Diederichs},\ and\ \citenamefont {Cassabois}}]{Nguyen2012}%
  \BibitemOpen
  \bibfield  {author} {\bibinfo {author} {\bibfnamefont {H.~S.}\ \bibnamefont
  {Nguyen}}, \bibinfo {author} {\bibfnamefont {G.}~\bibnamefont {Sallen}},
  \bibinfo {author} {\bibfnamefont {C.}~\bibnamefont {Voisin}}, \bibinfo
  {author} {\bibfnamefont {P.}~\bibnamefont {Roussignol}}, \bibinfo {author}
  {\bibfnamefont {C.}~\bibnamefont {Diederichs}}, \ and\ \bibinfo {author}
  {\bibfnamefont {G.}~\bibnamefont {Cassabois}},\ }\href {\doibase
  10.1103/PhysRevLett.108.057401} {\bibfield  {journal} {\bibinfo  {journal}
  {Phys. Rev. Lett.}\ }\textbf {\bibinfo {volume} {108}},\ \bibinfo {pages}
  {057401} (\bibinfo {year} {2012})}\BibitemShut {NoStop}%
\bibitem [{\citenamefont {Allen}\ and\ \citenamefont
  {Eberly}(1975)}]{Allen1975}%
  \BibitemOpen
  \bibfield  {author} {\bibinfo {author} {\bibfnamefont {L.}~\bibnamefont
  {Allen}}\ and\ \bibinfo {author} {\bibfnamefont {J.}~\bibnamefont {Eberly}},\
  }\href
  {http://books.google.com/books?hl=en&lr=&id=1q0ae-XNmWwC&oi=fnd&pg=PA1&dq=Optical+Resonance+and+Two-Level+Atoms&ots=mPT7_g7Wqt&sig=KNPpxRqSHF7aorRRsAI1mCOrIZ0}
  {\emph {\bibinfo {title} {{Optical resonance and two-level atoms}}}}\
  (\bibinfo  {publisher} {Dover},\ \bibinfo {year} {1975})\BibitemShut
  {NoStop}%
\bibitem [{\citenamefont {Flagg}\ \emph {et~al.}(2009)\citenamefont {Flagg},
  \citenamefont {Muller}, \citenamefont {Robertson}, \citenamefont {Founta},
  \citenamefont {Deppe}, \citenamefont {Xiao}, \citenamefont {Ma},
  \citenamefont {Salamo},\ and\ \citenamefont {Shih}}]{Flagg2009}%
  \BibitemOpen
  \bibfield  {author} {\bibinfo {author} {\bibfnamefont {E.~B.}\ \bibnamefont
  {Flagg}}, \bibinfo {author} {\bibfnamefont {a.}~\bibnamefont {Muller}},
  \bibinfo {author} {\bibfnamefont {J.~W.}\ \bibnamefont {Robertson}}, \bibinfo
  {author} {\bibfnamefont {S.}~\bibnamefont {Founta}}, \bibinfo {author}
  {\bibfnamefont {D.~G.}\ \bibnamefont {Deppe}}, \bibinfo {author}
  {\bibfnamefont {M.}~\bibnamefont {Xiao}}, \bibinfo {author} {\bibfnamefont
  {W.}~\bibnamefont {Ma}}, \bibinfo {author} {\bibfnamefont {G.~J.}\
  \bibnamefont {Salamo}}, \ and\ \bibinfo {author} {\bibfnamefont {C.~K.}\
  \bibnamefont {Shih}},\ }\href {\doibase 10.1038/nphys1184} {\bibfield
  {journal} {\bibinfo  {journal} {Nat. Phys.}\ }\textbf {\bibinfo {volume}
  {5}},\ \bibinfo {pages} {203} (\bibinfo {year} {2009})}\BibitemShut {NoStop}%
\bibitem [{\citenamefont {Georges}\ \emph {et~al.}(1979)\citenamefont
  {Georges}, \citenamefont {Lambropoulos},\ and\ \citenamefont
  {Zoller}}]{Georges1979}%
  \BibitemOpen
  \bibfield  {author} {\bibinfo {author} {\bibfnamefont {A.~T.}\ \bibnamefont
  {Georges}}, \bibinfo {author} {\bibfnamefont {P.}~\bibnamefont
  {Lambropoulos}}, \ and\ \bibinfo {author} {\bibfnamefont {P.}~\bibnamefont
  {Zoller}},\ }\href {\doibase 10.1103/PhysRevLett.42.1609} {\bibfield
  {journal} {\bibinfo  {journal} {Phys. Rev. Lett.}\ }\textbf {\bibinfo
  {volume} {42}},\ \bibinfo {pages} {1609} (\bibinfo {year}
  {1979})}\BibitemShut {NoStop}%
\bibitem [{\citenamefont {Rempe}\ \emph {et~al.}(1987)\citenamefont {Rempe},
  \citenamefont {Walther},\ and\ \citenamefont {Klein}}]{Rempe1987}%
  \BibitemOpen
  \bibfield  {author} {\bibinfo {author} {\bibfnamefont {G.}~\bibnamefont
  {Rempe}}, \bibinfo {author} {\bibfnamefont {H.}~\bibnamefont {Walther}}, \
  and\ \bibinfo {author} {\bibfnamefont {N.}~\bibnamefont {Klein}},\ }\href
  {\doibase 10.1103/PhysRevLett.58.353} {\bibfield  {journal} {\bibinfo
  {journal} {Phys. Rev. Lett.}\ }\textbf {\bibinfo {volume} {58}},\ \bibinfo
  {pages} {353} (\bibinfo {year} {1987})}\BibitemShut {NoStop}%
\bibitem [{\citenamefont {Mukamel}\ and\ \citenamefont
  {Dorfman}(2015)}]{Mukamel2015a}%
  \BibitemOpen
  \bibfield  {author} {\bibinfo {author} {\bibfnamefont {S.}~\bibnamefont
  {Mukamel}}\ and\ \bibinfo {author} {\bibfnamefont {K.~E.}\ \bibnamefont
  {Dorfman}},\ }\href {\doibase 10.1103/PhysRevA.91.053844} {\bibfield
  {journal} {\bibinfo  {journal} {Phys. Rev. A}\ }\textbf {\bibinfo {volume}
  {91}},\ \bibinfo {pages} {053844} (\bibinfo {year} {2015})}\BibitemShut
  {NoStop}%
\bibitem [{\citenamefont {Knight}\ and\ \citenamefont
  {Radmore}(1982)}]{Knight1982}%
  \BibitemOpen
  \bibfield  {author} {\bibinfo {author} {\bibfnamefont {P.}~\bibnamefont
  {Knight}}\ and\ \bibinfo {author} {\bibfnamefont {P.}~\bibnamefont
  {Radmore}},\ }\href {\doibase 10.1016/0375-9601(82)90625-9} {\bibfield
  {journal} {\bibinfo  {journal} {Phys. Lett. A}\ }\textbf {\bibinfo {volume}
  {90}},\ \bibinfo {pages} {342} (\bibinfo {year} {1982})}\BibitemShut
  {NoStop}%
\bibitem [{\citenamefont {Ramsay}\ \emph {et~al.}(2010)\citenamefont {Ramsay},
  \citenamefont {Gopal}, \citenamefont {Gauger}, \citenamefont {Nazir},
  \citenamefont {Lovett}, \citenamefont {Fox},\ and\ \citenamefont
  {Skolnick}}]{Ramsay2010}%
  \BibitemOpen
  \bibfield  {author} {\bibinfo {author} {\bibfnamefont {A.~J.}\ \bibnamefont
  {Ramsay}}, \bibinfo {author} {\bibfnamefont {A.~V.}\ \bibnamefont {Gopal}},
  \bibinfo {author} {\bibfnamefont {E.~M.}\ \bibnamefont {Gauger}}, \bibinfo
  {author} {\bibfnamefont {A.}~\bibnamefont {Nazir}}, \bibinfo {author}
  {\bibfnamefont {B.~W.}\ \bibnamefont {Lovett}}, \bibinfo {author}
  {\bibfnamefont {A.~M.}\ \bibnamefont {Fox}}, \ and\ \bibinfo {author}
  {\bibfnamefont {M.~S.}\ \bibnamefont {Skolnick}},\ }\href {\doibase
  10.1103/PhysRevLett.104.017402} {\bibfield  {journal} {\bibinfo  {journal}
  {Phys. Rev. Lett.}\ }\textbf {\bibinfo {volume} {104}},\ \bibinfo {pages}
  {017402} (\bibinfo {year} {2010})}\BibitemShut {NoStop}%
\bibitem [{\citenamefont {Kuhlmann}\ \emph {et~al.}(2015)\citenamefont
  {Kuhlmann}, \citenamefont {Prechtel}, \citenamefont {Houel}, \citenamefont
  {Ludwig}, \citenamefont {Reuter}, \citenamefont {Wieck},\ and\ \citenamefont
  {Warburton}}]{Kuhlmann2015}%
  \BibitemOpen
  \bibfield  {author} {\bibinfo {author} {\bibfnamefont {A.~V.}\ \bibnamefont
  {Kuhlmann}}, \bibinfo {author} {\bibfnamefont {J.~H.}\ \bibnamefont
  {Prechtel}}, \bibinfo {author} {\bibfnamefont {J.}~\bibnamefont {Houel}},
  \bibinfo {author} {\bibfnamefont {A.}~\bibnamefont {Ludwig}}, \bibinfo
  {author} {\bibfnamefont {D.}~\bibnamefont {Reuter}}, \bibinfo {author}
  {\bibfnamefont {A.~D.}\ \bibnamefont {Wieck}}, \ and\ \bibinfo {author}
  {\bibfnamefont {R.~J.}\ \bibnamefont {Warburton}},\ }\href {\doibase
  10.1038/ncomms9204} {\bibfield  {journal} {\bibinfo  {journal} {Nat.
  Commun.}\ }\textbf {\bibinfo {volume} {6}},\ \bibinfo {pages} {8204}
  (\bibinfo {year} {2015})}\BibitemShut {NoStop}%
\bibitem [{\citenamefont {Cohen-Tannoudji}\ and\ \citenamefont
  {Reynaud}(1977)}]{Cohen-Tannoudji1977}%
  \BibitemOpen
  \bibfield  {author} {\bibinfo {author} {\bibfnamefont {C.}~\bibnamefont
  {Cohen-Tannoudji}}\ and\ \bibinfo {author} {\bibfnamefont {S.}~\bibnamefont
  {Reynaud}},\ }\href {\doibase 10.1088/0022-3700/10/3/005} {\bibfield
  {journal} {\bibinfo  {journal} {J. Phys. B At. Mol. Phys.}\ }\textbf
  {\bibinfo {volume} {10}},\ \bibinfo {pages} {345} (\bibinfo {year}
  {1977})}\BibitemShut {NoStop}%
\bibitem [{\citenamefont {Avan}\ and\ \citenamefont
  {Cohen-Tannoudji}(1977)}]{Avan1977}%
  \BibitemOpen
  \bibfield  {author} {\bibinfo {author} {\bibfnamefont {P.}~\bibnamefont
  {Avan}}\ and\ \bibinfo {author} {\bibfnamefont {C.}~\bibnamefont
  {Cohen-Tannoudji}},\ }\href {\doibase 10.1088/0022-3700/10/2/006} {\bibfield
  {journal} {\bibinfo  {journal} {J. Phys. B}\ }\textbf {\bibinfo {volume}
  {155}},\ \bibinfo {pages} {117} (\bibinfo {year} {1977})}\BibitemShut
  {NoStop}%
\bibitem [{\citenamefont {Matthiesen}\ \emph {et~al.}(2012)\citenamefont
  {Matthiesen}, \citenamefont {Vamivakas},\ and\ \citenamefont
  {Atat{\"{u}}re}}]{Matthiesen2012}%
  \BibitemOpen
  \bibfield  {author} {\bibinfo {author} {\bibfnamefont {C.}~\bibnamefont
  {Matthiesen}}, \bibinfo {author} {\bibfnamefont {A.~N.}\ \bibnamefont
  {Vamivakas}}, \ and\ \bibinfo {author} {\bibfnamefont {M.}~\bibnamefont
  {Atat{\"{u}}re}},\ }\href {\doibase 10.1103/PhysRevLett.108.093602}
  {\bibfield  {journal} {\bibinfo  {journal} {Phys. Rev. Lett.}\ }\textbf
  {\bibinfo {volume} {108}},\ \bibinfo {pages} {093602} (\bibinfo {year}
  {2012})}\BibitemShut {NoStop}%
\bibitem [{\citenamefont {Ulrich}\ \emph {et~al.}(2011)\citenamefont {Ulrich},
  \citenamefont {Ates}, \citenamefont {Reitzenstein}, \citenamefont
  {L{\"{o}}ffler}, \citenamefont {Forchel},\ and\ \citenamefont
  {Michler}}]{Ulrich2011}%
  \BibitemOpen
  \bibfield  {author} {\bibinfo {author} {\bibfnamefont {S.~M.}\ \bibnamefont
  {Ulrich}}, \bibinfo {author} {\bibfnamefont {S.}~\bibnamefont {Ates}},
  \bibinfo {author} {\bibfnamefont {S.}~\bibnamefont {Reitzenstein}}, \bibinfo
  {author} {\bibfnamefont {A.}~\bibnamefont {L{\"{o}}ffler}}, \bibinfo {author}
  {\bibfnamefont {A.}~\bibnamefont {Forchel}}, \ and\ \bibinfo {author}
  {\bibfnamefont {P.}~\bibnamefont {Michler}},\ }\href {\doibase
  10.1103/PhysRevLett.106.247402} {\bibfield  {journal} {\bibinfo  {journal}
  {Phys. Rev. Lett.}\ }\textbf {\bibinfo {volume} {106}},\ \bibinfo {pages}
  {247402} (\bibinfo {year} {2011})}\BibitemShut {NoStop}%
\bibitem [{\citenamefont {Ulhaq}\ \emph {et~al.}(2012)\citenamefont {Ulhaq},
  \citenamefont {Weiler}, \citenamefont {Ulrich}, \citenamefont {Ro{\ss}bach},
  \citenamefont {Jetter},\ and\ \citenamefont {Michler}}]{Ulhaq2012}%
  \BibitemOpen
  \bibfield  {author} {\bibinfo {author} {\bibfnamefont {A.}~\bibnamefont
  {Ulhaq}}, \bibinfo {author} {\bibfnamefont {S.}~\bibnamefont {Weiler}},
  \bibinfo {author} {\bibfnamefont {S.~M.}\ \bibnamefont {Ulrich}}, \bibinfo
  {author} {\bibfnamefont {R.}~\bibnamefont {Ro{\ss}bach}}, \bibinfo {author}
  {\bibfnamefont {M.}~\bibnamefont {Jetter}}, \ and\ \bibinfo {author}
  {\bibfnamefont {P.}~\bibnamefont {Michler}},\ }\href {\doibase
  10.1038/nphoton.2012.23} {\bibfield  {journal} {\bibinfo  {journal} {Nat.
  Photonics}\ }\textbf {\bibinfo {volume} {6}},\ \bibinfo {pages} {238}
  (\bibinfo {year} {2012})}\BibitemShut {NoStop}%
\bibitem [{\citenamefont {Schubert}\ \emph {et~al.}(1979)\citenamefont
  {Schubert}, \citenamefont {S{\"{u}}sse},\ and\ \citenamefont
  {Vogel}}]{Schubert1979}%
  \BibitemOpen
  \bibfield  {author} {\bibinfo {author} {\bibfnamefont {M.}~\bibnamefont
  {Schubert}}, \bibinfo {author} {\bibfnamefont {K.-E.}\ \bibnamefont
  {S{\"{u}}sse}}, \ and\ \bibinfo {author} {\bibfnamefont {W.}~\bibnamefont
  {Vogel}},\ }\href {\doibase 10.1016/0030-4018(79)90352-3} {\bibfield
  {journal} {\bibinfo  {journal} {Opt. Commun.}\ }\textbf {\bibinfo {volume}
  {30}},\ \bibinfo {pages} {275} (\bibinfo {year} {1979})}\BibitemShut
  {NoStop}%
\end{thebibliography}%
\end{document}